\documentclass[lettersize,journal]{IEEEtran}
\usepackage{amsmath,amsfonts}
\usepackage{algorithmic}
\usepackage{algorithm}
\usepackage{array}
\usepackage{textcomp}
\usepackage{stfloats}
\usepackage{url}
\usepackage{verbatim}
\usepackage{graphicx}
\usepackage{cite}
\usepackage{color}
\usepackage{caption}
\usepackage{subcaption}
\usepackage{cite}

\begin{document}

\title{
Asymmetric Modular Pulse Synthesizer: A High-Power High-Granularity Electronics Solution for Transcranial Magnetic Stimulation with Practically Any Pulse Shape for Neural Activation Selectivity
}

\author{
	\vskip 1em
	Jinshui Zhang, 
	Angel V.~Peterchev, 
	Stefan M.~Goetz
    }

\maketitle

\begin{abstract}
    Noninvasive brain stimulation can activate neurons in the brain but requires power electronics with exceptionally high power in the mega-volt-ampere  and high frequencies in the kilohertz range. Whereas oscillator circuits offered only one or very few pulse shapes, modular power electronics solved a long-standing problem for the first time and enabled arbitrary software-based design of the temporal shape of stimuli. However, synthesizing arbitrary stimuli with a high output quality requires a large number of modules. Systems with few modules and pulse-width modulation may generate apparently smooth current shapes in the highly inductive coil, but the stimulation effect of the neurons depends on the electric field and the electric field becomes a burst of ultra-brief rectangular pulses.
    
    We propose an alternative solution that achieves high-resolution pulse shaping with fewer modules by implementing high-power wide-bandwidth voltage asymmetry. Rather than  equal voltage steps, our system strategically assigns different voltages to each module to achieve a near-exponential improvement in resolution. Compared to prior designs, our experimental prototype achieved better output quality, although it uses only half the number of modules.
\end{abstract}


Transcranial magnetic stimulation (TMS) can noninvasively write signals into neurons in the brain and modulate networks with certain pulse rhythms \cite{hallett2007transcranial,brignani2008modulation}. Strong brief current pulses in the stimulation coil generate time-varying magnetic fields, which reach into the body and induce electric fields. The electric fields in turn can activate electrically sensitive ion channels, nanomechanics in the form of transmembrane proteins, which change their shape in an electric field \cite{KLOMJAI2015208,armstrong1998voltage}. The selective activation of neural pathways and neuromodulation with TMS are widely used in experimental brain research and clinical therapy \cite{LEFAUCHEUR2020474,hallett2007transcranial}. However, even with the most focal coils, TMS still activates a large number of neurons in the focus of the coil with each pulse. Furthermore, neuromodulation is often variable and potentially a net imbalance between simultaneous inhibitory and excitatory effects \cite{KLOMJAI2015208}.

Whereas noninvasive coil technology has physical limits and does not promise to ever achieve cell-level focality, a growing body of literature reports that the temporal shape of a pulse, e.g., monophasic, biphasic, and polyphasic in the oldest devices, can exploit the different activation dynamics of various neural elements \cite{gomez2018design, pisoni2018tms, sommer2014tms, halawa2019neuronal,  KAMMER2001250, kammer2007anisotropy, ROTHKEGEL20101915, sommer2006half, kammer2001influence,lazzaro2001comparison,rothkegel2010impact,cadwell1991optimizing} and also the neuromodulation strength and effect depends on the pulse shape \cite{ARAI2005605, taylor2007stimulus,goetz2016enhancement, HANNAH2017106,hosono2008comparison,antal2002pulse,sommer2002neuronal,arai2005differences}.
However, due to the high power of pulses and relatively high frequency content, TMS device technology is very limited.

\begin{figure*}
    \centering
    \includegraphics{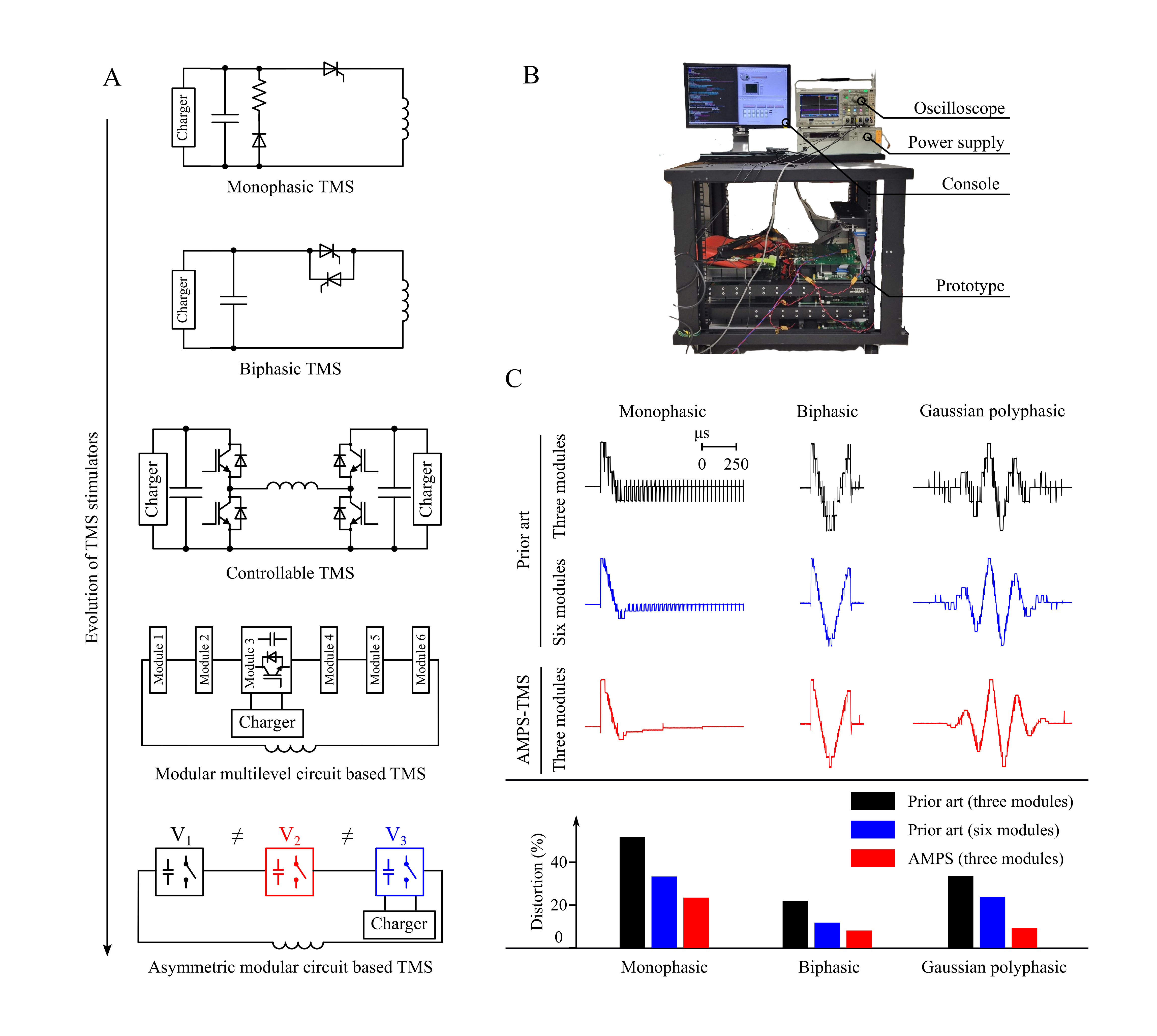}
    \caption{Asymmetric modular pulse synthesizer (AMPS) TMS demonstration.
	(A) Typical TMS setup, consisting of a stimulator, a coil, and auxiliary components such as monitoring systems and control consoles.
	(B) Evolution of TMS stimulator circuits, progressing from conventional monophasic and biphasic oscillators to controllable TMS, and further to modular multilevel circuit based TMS.
	(C) Experimental prototype and testing platform for the proposed AMPS-TMS technology.
	(D) Experimental results showcasing monophasic, biphasic, and Gaussian polyphasic waveforms, compared to conventional multilevel TMS stimulators.
    }
    \label{fig:main_letter}
\end{figure*}

Progress in electronics has substantially enhanced stimulator technology (see Fig.\ 1B for evolution). The earliest generation used oscillator circuits, but each device could only produce one fixed pulse type---specifically monophasic and biphasic, whereas polyphasic and half-wave are rare---with fixed durations \cite{goetz2017development}. The oscillator allows to slowly accumulate energy up to several hundred joules during charging and release it in a very brief interval, e.g., several hundred microseconds. However, the oscillators limit the pulse shape and duration according to the electrical capacitance of the internal capacitor and the magnetic inductance of the treatment coil.

Various approaches either reconfigured the oscillator before a pulse to change the pulse shape between a few different options, or during a pulse to shape the pulse. Controllable pulse parameter TMS (cTMS) generates a near-rectangular pulse shape 
and forms a rather large oscillator, of which it actively terminates a phase with a powerful transistor. This concept has been revisited by various other practices \cite{4360051,Peterchev_2011,6186791}. Although cTMS provides some controllability of the pulse, it is still bound to certain classes of pulse shapes associated with the implemented oscillators. These pulses are rectangular, which may have energy advantages \cite{goetz2013analysis, ma2023optimisation, Wang_2023, 6347004}, but cTMS cannot generate conventional monophasic or biphasic pulses with their sinusoidal segments. Greater flexibility is needed to achieve original sinusoidal waveforms or even entirely user-designed shapes.

Latest technologies allow synthesizing any practical pulse shape, including conventional monophasic and biphasic pulses, and turned pulse generation into a software problem \cite{goetz2012circuit,Li_2022, Zeng_2022}. These modular systems split the overall voltage and power of TMS pulses into smaller, manageable units and generate the required voltage at each time by combining the outputs of individual modules. Each module contributes a step to the voltage and the electric field. The overall granularity of pulse generation grows with the number of modules. Fast switching between steps, so-called modulation, can increase fine control to some extent. However, very fine shaping of the output and low deviation still requires many modules.

Systems with fewer modules strongly rely on switching modulation, e.g., pulse-width modulation as known from motor drives and similar inverter applications \cite{reimers2019automotive, kamiyama1989pwm, Sinisalo2021}. However, approximating TMS pulses by rapidly switching the voltage on and off or between only few levels was found to generate detectable deviations compared to conventional pulses \cite{MEMARIANSORKHABI2022980, SINISALO2024218}. However, in contrast to motor drives, where the current matters, the stimulating effect of TMS relies on the induced electric field: Although the coil current may appear passably smooth and presentable, the electric field rather turns  into a burst of ultra-brief high-amplitude rectangular pulses, each with microsecond duration. Figure 1C illustrates how course pulses with conventional pulse-width modulation with three modules become, whereas the literature even suggests fewer modules and therefore levels.

The early understanding suggested that neurons are slow in response and potentially in the first approximation even linear so that it does not matter. However, recent work with high-frequency content indicates that this is wrong, and the deviations in motor threshold to the pulses that should be approximated corroborate that.
Neurons are known to exhibit strongly nonlinear behavior so that a linear separation into different spectral contents, which are supposed to act on neurons without interaction, does not work \cite{victor1977nonlinear}.

We therefore suggest  an alternative. We use fewer modules but still generate fairly smooth electric field pulse shapes with high granularity. Instead of equal voltage steps, we designed a system where we operate the modules at different voltages. This trick allows to dramatically refine the electric field granularity already with a small number of modules: whereas for the previous TMS technology with free pulse shape design, the electric-field granularity increased linearly with the number of modules, our output granularity grows now nearly exponentially. 

With unequal voltages in the modules, we can also generate all sum and difference voltage levels, which increases the options manifold with every additional module. 
We do not use any obvious module voltage sequence such as binary or ternary due to the huge voltage steps from one module to the next. Such large voltage steps are very impractical as they require dedicated module designs for every level and further again a larger number of modules or higher voltage per module to reach TMS levels.
Instead, we present a three-module prototype of an asymmetric modular pulse synthesizer with a maximum 50\% voltage difference from module to module, corresponding to a 1.5 voltage differential ratio between adjacent modules.
The granularity with which the electric field can be shaped grows drastically. 

Figure 1C represents our experimental prototype. The system is expandable with at present 1.5 kV, by adding more of the same circuit boards. We set up devices and recorded pulse shapes from three systems: one conventional modular pulse synthesizer with three modules and one with six modules as well as an asymmetric device as suggested here.
The recordings in Fig.\ 1D compare the output of the new prototype with the older technology across various pulse shapes: monophasic, biphasic, and a polyphasic pulse with Gaussian envelop. 

Despite only three modules, our prototype can generate 27 voltage and electric field steps. It can rapidly change in between to assemble a smooth electric field profile of practically any shape. The consequence is a visibly and also quantitatively low distortion. Our prototype consistently outperforms all variations of prior asymmetric modular circuits from other fields of electronics. It demonstrates significantly lower distortion levels for all pulse shapes tested. This advantage is particularly evident for pulses that include long, shallow electric field phases, such as monophasic and Gaussian polyphasic.

In conclusion, incorporating asymmetry into modular TMS circuits can significantly enhance the output granularity. This work demonstrates that AMPS-TMS machines can achieve high output quality with reduced switching rates and fewer modules. This technology promises to lower the cost of flexible TMS stimulators, as it requires fewer modules and re-opens the possibility of using slower but more affordable transistors, such as insulated-gate bipolar transistors (IGBT). We further propose that tailoring asymmetric voltage profiles, such as the 1.5 ratio used in this excercise, can improve the practicality of asymmetric modular TMS circuits with high pulse quality.

\bibliographystyle{Bibliography/IEEEtranTIE}
\bibliography{Bibliography/IEEEabrv,Bibliography/main}\ 

\end{document}